\documentclass[twocolumn,showpacs,floatfix,pre]{revtex4}

\usepackage{amssymb,amsmath,epsfig,graphicx}

\usepackage[FIGTOPCAP]{subfigure}

\newcommand{\Feff}{F_{\text{eff}}}
\newcommand{\Veff}{V_{\text{eff}}}
\newcommand{\barFeff}{\bar F_{\text{eff}}}
\newcommand{\barVeff}{\bar V_{\text{eff}}}

\begin{document}

\title{Rocking feedback-controlled ratchets}

\author{M. Feito}
\email{feito@fis.ucm.es}
\affiliation{Departamento de F\'{\i}sica At\'omica, Molecular y
Nuclear, Universidad Complutense de Madrid, \\
Avenida Complutense s/n, 28040 Madrid, Spain.}

\author{J. P. Baltan\'as}
\email{baltanas@us.es}
\affiliation{Departamento de F\'{\i}sica Aplicada II, Universidad de Sevilla,\\
Av. Reina Mercedes 2, 41012 Sevilla, Spain.}

\author{F. J. Cao}
\email{francao@fis.ucm.es}
\affiliation{Departamento de F\'{\i}sica At\'omica, Molecular y
Nuclear, Universidad Complutense de Madrid, \\
Avenida Complutense s/n, 28040 Madrid, Spain,}
\affiliation{LERMA, Observatoire de Paris,
Laboratoire Associ\'e au CNRS UMR 811 2,\\
61, Avenue de l'Observatoire, 75014 Paris, France.}


\begin{abstract}
We investigate the different regimes that emerge when a periodic driving
force, the rocking force, acts on a collective feedback flashing ratchet.
The interplay of the rocking and the feedback control gives a rich dynamics
with different regimes presenting several unexpected
features. In particular, we show that for both the one-particle ratchet
and the collective version of the ratchet an appropriate rocking increases the
flux. This mechanism gives the maximum flux that has been achieved in
a ratchet device without an a priori bias.
\end{abstract}



\maketitle


\section{Introduction}

Ratchets can be viewed as controllers that act on
stochastic systems with the aim of inducing directed motion by breaking of
thermal equilibrium and certain time-space symmetries~\cite{rei02}. As
usual in control theory~\cite{bec05}, these systems are divided into
\emph{open-loop} ratchets~\cite{rei02}, when the actuation does not use any
knowledge of the state of the system; and \emph{closed-loop}
ratchets~\cite{cao04,cra08}, when information on the state of the system
is used to decide how to operate on the system. These closed-loop ratchets
---also called feedback or information ratchets--- have recently
attracted attention as Maxwell's demon devices that are capable of maximizing
the performance of ratchets~\cite{fei07}. They may also be
relevant to get insight into the motion of linear, two headed, processive
molecular motors~\cite{bie07}. In addition, experimental realizations of
feedback ratchets have been recently proposed~\cite{cra08,fei07b} and
implemented~\cite{lop08} due to their potential relevance as nanotechnological
devices.
\par

A relevant class of ratchets are \emph{flashing} ratchets, which
operate switching on and off a spatially periodic potential.
Flashing ratchets have been studied in both open-loop
(e.g.~\cite{rei02,bug87}) and closed-loop
(e.g.~\cite{cao04,fei06,cra08}) schemes. A generalization of these
ratchets are pulsated ratchets~\cite{rei02}, in which the amplitude of the
ratchet potential is modulated in time, but not necessarily flashed on and off.
On the other hand, \emph{rocking} ratchets operate thanks to a periodic driving
force, and thus they perform an open-loop control. Rocking ratchets reveals a
rich dynamics, which includes current reversals, distinct
stable trajectories, and quantization of the deterministic
current~\cite{bar94,mag93,ajd94}. The combination of open-loop
pulsated ratchets and rocking ratchets has been studied in
Refs.~\cite{sav04}, giving the possibility of a reverse of
the sign of the flux with respect to the simple rocked ratchet.
\par

In the present paper we study the effects of adding a periodic
driving force that rocks a feedback-controlled flashing ratchet.
We analyze the new intriguing dynamics that emerge due to the
interplay between the feedback control and the rocking. In
particular, we show that the rocking of a feedback ratchet allows
the system to improve the flux performance. The optimization of
the flux performance of ratchets is potentially relevant for their
nanotechnological applications, and the enhancement of the flux
performance in flashing ratchets due to feedback has
been recently verified experimentally~\cite{lop08}. We show here
how this flux performance can be further improved thanks to the
effects produced by an additional rocking force. In the next
section we describe the rocked feedback-controlled ratchet, and
after, in Sec.~\ref{sec:one}, we study the one-particle ratchet.
The collective version of the ratchet compounded of more than one
particle is analyzed in  Sec.~\ref{sec:collective} in the regimes
of few and many particles. We finally summarize and comment our
main results in Sec.~\ref{sec:conclusions}.

\section{Rocked feedback-controlled ratchet}\label{sec:model}
Let us consider $N$ Brownian particles at temperature $T$ in a periodic
potential $V(x)$, the ratchet potential. The state of the system is described
by the positions $x_i(t)$ of the particles ($i=1,\dots,N$) satisfying the
overdamped Langevin equations with a fluctuating (rocking) force of amplitude
$A$ and frequency $\Omega$,
\begin{equation}\label{langevin}
\gamma \dot x_i(t)=\alpha(x_1(t),\dots,x_N(t),t)F(x_i(t))+A\cos (\Omega
t)+\xi_i(t).
\end{equation}
Here, $F(x)=-V'(x)$, $\gamma$ is the friction coefficient (related to the
diffusion coefficient $D$ through Einstein's relation
$D=k_BT/\gamma$), and $\xi_i(t)$ are Gaussian white noises of zero
mean and variance $\langle \xi_i(t)\xi_j(t^\prime)\rangle =2\gamma
k_B T\delta_{ij}\delta(t-t^\prime)$.
Note that the control parameter $\alpha$ depends explicitly on the state of
the system. Therefore, this ratchet is feedback controlled, what implies
an effective coupling between the particles.
\par

In order to quantify the inducted directed motion a relevant
quantity is the stationary center-of-mass velocity or flux defined
as
\begin{equation}\label{last}
\langle \dot x\rangle := \lim_{t\to\infty}
\frac{1}{N}\sum_{i=1}^N\frac{\langle x_i(t) - x_i(0) \rangle}{t} =
\lim_{t\to\infty} \frac{1}{N}\sum_{i=1}^N\frac{\langle
x_i(t)\rangle}{t}.
\end{equation}
Due to the long-time limit and the average over realizations this
asymptotic center-of-mass velocity does not depend neither on the
phase of the fluctuating force nor on the initial particle
positions~\cite{rei02}. We shall consider the relevant control
policy that maximizes the instant center-of-mass velocity
introduced in~\cite{cao04}. In this feedback protocol, the
controller computes the force per particle due to the ratchet
potential if it were on,
\begin{equation}\label{netforce}
f(x_1(t),\dots,x_N(t))=\frac{1}{N}\sum_{i=1}^N F(x_i(t)),
\end{equation}
and switches the potential on ($\alpha=1$) if $f(t)$ is
positive or switches the potential off ($\alpha=0$) otherwise. Therefore, the
feedback control protocol considered is
\begin{equation}\label{alfa}
  \alpha(x_1(t),\dots,x_N(t))= \Theta(f(x_1(t),\dots,x_N(t))),
\end{equation}
with $\Theta$ the Heaviside function [$\Theta (x)=1$ if $x>0$,
else $\Theta (x)=0$].
\par

\begin{figure}
  \begin{center}
    \includegraphics [scale=0.6] {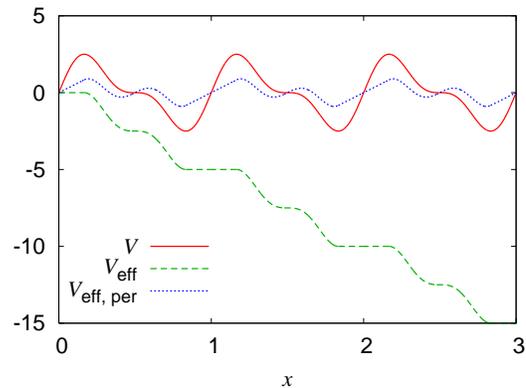}
  \end{center}
  \caption{(Color online) Ratchet potential $V$ (solid red line)
    [Eq.~\eqref{smoothpot} with $V_0 = 5 $], one-particle effective
  potential $\Veff$ (dashed green line), and  one-particle periodic effective
  potential $\Veff^{\text{per}}$ (dotted blue line). Units: $ k_B T = 1 $,
  $ L = 1 $.
  }
  \label{fig:pot}
\end{figure}

The graphs that illustrate the results of this paper have been obtained
considering the periodic asymmetric potential
\begin{equation}\label{smoothpot}
V(x) = \frac{2V_0}{3\sqrt{3}} \left[ \sin \left(\frac{2\pi x}{L}
\right) + \frac{1}{2} \sin \left( \frac{4\pi x}{L} \right)
\right],
\end{equation}
which has potential height $V_0$ and period $L$; see Fig.~\ref{fig:pot}.
We can introduce an asymmetry parameter $a$ for the potential such that $aL$
is defined as the distance between a minimum of the
potential and the first maximum at the righthand-side. The
potential in Eq.~\eqref{smoothpot} has an asymmetry parameter of
$a=1/3$. We have also performed computations with
other potentials and found analogous results.
\par

In order to numerically compute the flux~\eqref{last} we have
performed numerical simulations of the Langevin
equations~\eqref{langevin} [with the control parameter $\alpha$
given by Eq.~\eqref{alfa}] by using the Euler-Maruyama
algorithm~\cite{klo92}. We have verified that our numerical
results were not affected by systematic errors due to time
discretization, initial transitories or finite number of
realizations. The characteristic time that takes the system to
diffuse the distance $aL$ of the uphill part of the ratchet
potential is $a^2L^2/(2D)$, while the characteristic time that
takes to `slide down' the downhill part of the potential is
$\gamma(1-a)^2L^2/V_0$. Therefore, the discretization time has
been chosen much smaller than these times, and also much smaller
than the period of the rocking, $2\pi/\Omega$, to avoid aliasing.
In addition, the Langevin equations have been numerically solved
up to times much greater than these characteristic times, and
large enough to ensure a value of the computed flux close to its
asymptotic value independently of the specific realization of the
stochastic process, the initial conditions, or the initial
transitories. We have performed the number of realizations
required to have small statistical errors in the averages
computed.
\par

\section{One-particle ratchet}\label{sec:one}
For the one-particle ratchet ($N=1$), the maximization of the instant velocity
control policy, Eq.~\eqref{alfa}, only depends
of the position $x(t)$ of the particle. Hence, we can define
an effective force $\Feff(x)=\alpha(x) F(x)$ that allows us to rewrite the
Langevin equation~\eqref{langevin} as
\begin{equation}\label{langevinoneFeeedbackF}
\gamma\dot x(t)=\Feff(x(t)) + A\cos(\Omega t) + \xi(t).
\end{equation}
The effective force $\Feff$ derives from an effective potential
$\Veff(x)$ that is no longer periodic, but tilted downhill. This
$\Veff$ can be recast as a periodic potential
$V_{\text{eff}}^\text{per}(x)$ of height $aV_0$ and asymmetry $a$,
plus a linear term $ V_0 x / L $ accounting for the bias, where $
V_0 $ is the height of the ratchet potential, $a$ its asymmetry
parameter, and $ L $ its period. Therefore, we can write $
\Veff(x) = \Veff^{\text{per}}(x) - V_0 x / L $, as we illustrate
in Fig.~\ref{fig:pot} for potential~\eqref{smoothpot}. In view
of these considerations, the feedback rocking ratchet can be
reinterpreted as an open-loop rocking ratchet with a biased
asymmetric potential. Thus Eq.~\eqref{langevinoneFeeedbackF}
stands for the celebrated SQUID ratchet~\cite{zap96},
\begin{equation} \label{SQUID}
\gamma\dot x(t)=-\frac{d}{dx}\Veff^{\text{per}}(x(t))+ V_0/L + A\cos(\Omega t) +
\xi(t).
\end{equation}
This equation of motion describes the dynamics of a tilted rocking
ratchet, i.e., of a periodically driven single Brownian particle in
a tilted washboard potential, and it has been extensively studied
analytically and numerically
\cite{zap96,ajd94,reg02} (even when inertial
terms are also present~\cite{mat08}). For
instance, for the adiabatic regime, i.e., the regime of slow
driving~\cite{mag93},  the flux can be approximated by
\begin{equation} \label{adiabatic}
\langle\dot{x}\rangle= \frac{1}{\cal T}\int_{0}^{\cal T}
\langle\dot{x}\rangle_{G(t)}\;dt,
\end{equation}
where ${\cal T}=2\pi/\Omega$ is the period of the driving force
and $\langle\dot{x}\rangle_{G(t)}$ is the asymptotic flux that
would be obtained if the driving force were fixed at the instant
$t$ to its value $G(t)=A\cos(\Omega t)$. This flux can be obtained
by solving a Fokker-Plank equation for the resulting constant
external force~\cite{fei07}. Other analytical results have been
reported for the high-frequency regime~\cite{pla98}, or the
deterministic (zero-temperature) regime~\cite{ben02}. Thanks
to the equivalence found between the one-particle rocked feedback
ratchet [Eqs.~\eqref{langevin} and \eqref{alfa}] and the SQUID
ratchet [Eq.~\eqref{SQUID}], all the effects found for the rocked
feedback ratchet have their counterparts in the extensively
studied SQUID ratchet. However, it is important to emphasize that
the tilt appears in our results not as an a priori bias, but as part
of an effective description of the effects of the feedback.
\par

Here, we discuss the results for the different regimes obtained by
performing numerical simulations of the Langevin equation. We
shall first discuss the case of zero temperature and later the
case of nonzero temperature.
\par

\begin{figure}
  \begin{center}
    \includegraphics [scale=0.6] {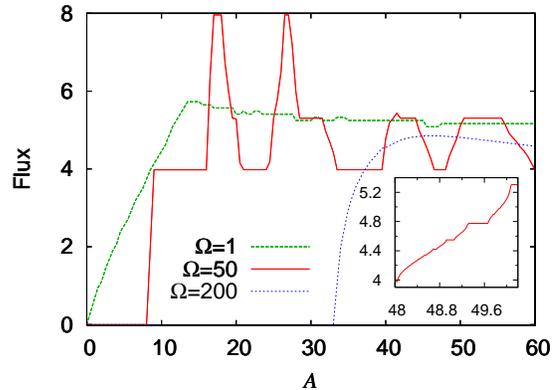}
  \end{center}
  \caption{(Color online) One-particle case. Flux for the deterministic
    (zero temperature) rocked feedback ratchet as a function of the amplitude
    $A$ of the rocking and frequencies $\Omega=1, 50, 200$ for the ratchet
    potential of Fig.~\ref{fig:pot}. {\em Inset:} Zoom of the flux for
    $\Omega=50$ and $A$ belonging to the interval $[48.0,50.2]$. Units: $V_0=5$,
    $L=1$, $\gamma=1$.}\label{fig:quant}
\end{figure}

In the deterministic case of zero temperature there is no
diffusion and only the rocking force can help the particle to
cross the flat regions of the effective potential $ \Veff $ (see
Fig.~\ref{fig:pot}). This makes the  flux strictly zero for small
amplitudes such that the particle cannot overcome the flat part of
the effective tilted potential; see Fig.~\ref{fig:quant}. For
higher amplitudes the flux exhibits remarkable characteristic
effects. Our simulations show that the deterministic flux is
quantized and it presents a step-like structure, a well-known
effect for open-loop rocking ratchets~\cite{zap96}. This structure
is specially clear for the frequency $\Omega=50$ in
Fig.~\ref{fig:quant}. The flux quantization is owing to the
synchronization with the phase of the periodic driving
(see~\cite{zap96} for details). Its step-like structure presents a
self-similar structure with steps at rational values of the flux,
which can be seen performing successive zoom-in views (see inset
in Fig. 2). This structure is known as Devil's
staircase~\cite{ajd94,zap96}.
\par

\begin{figure}
  \begin{center}
    \subfigure[]{\includegraphics[scale=0.6]{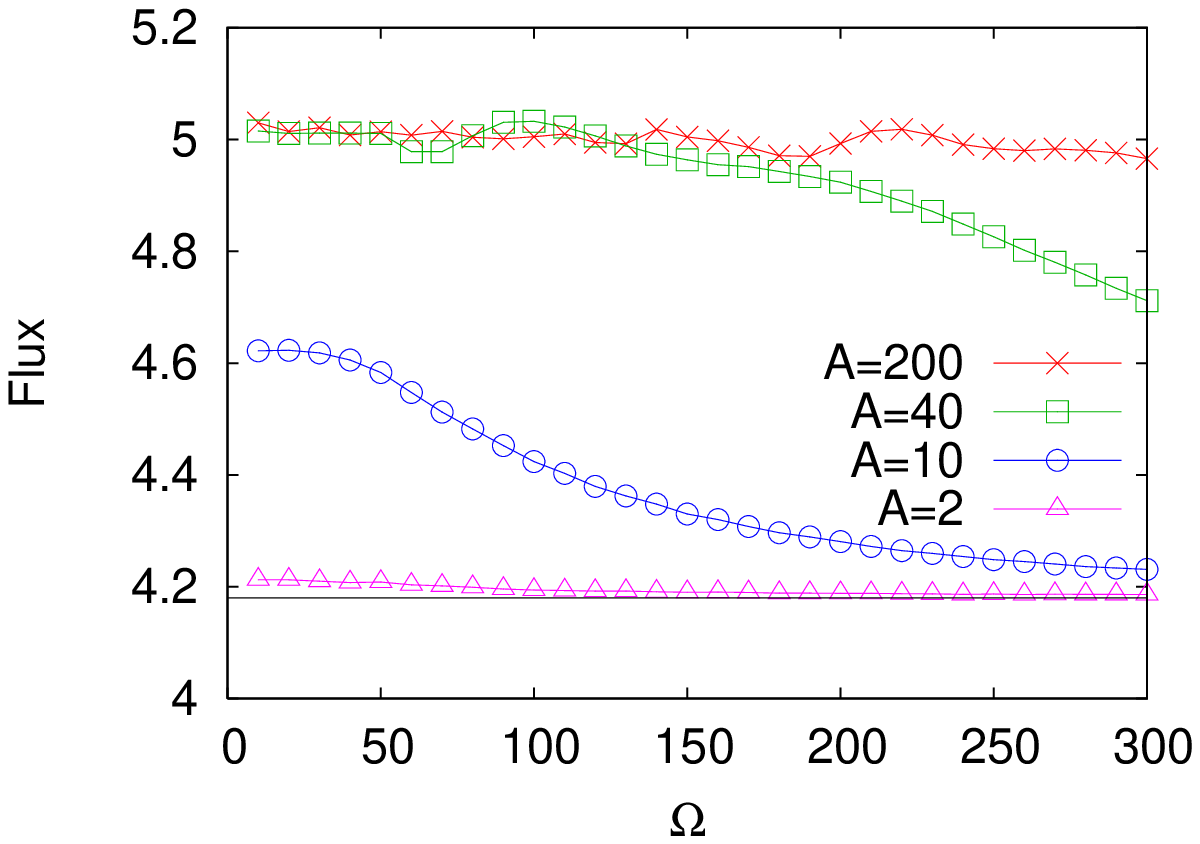}}
    \subfigure[]{\includegraphics[scale=0.6]{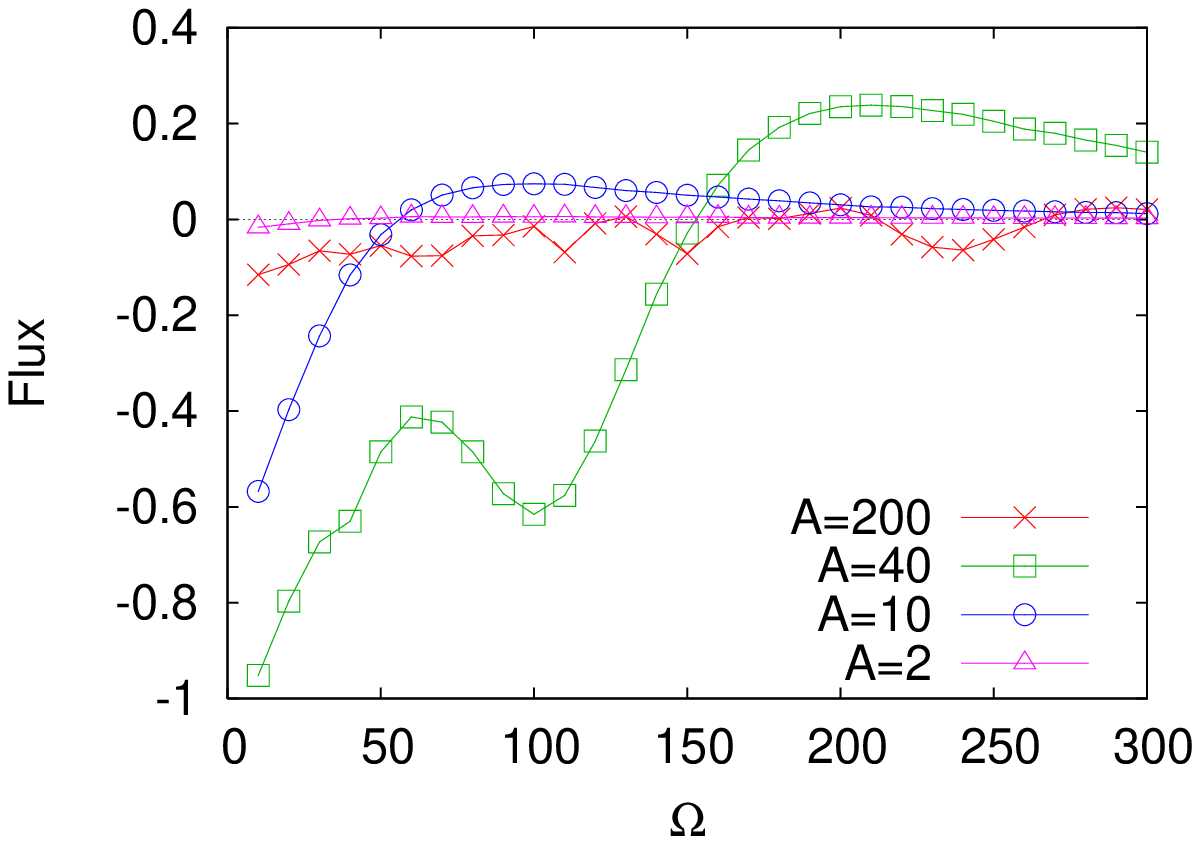}}
\end{center}
\caption{(Color online) One-particle case.
    \emph{Panel (a):} Flux $ \langle \dot x \rangle $ versus frequency
    $\Omega$ for the rocked
    feedback ratchet. The horizontal solid line stands for the pure feedback
    ratchet ---without rocking, i.e., $A=0$.---
    \emph{Panel (b):} Flux $ \langle \dot x \rangle $
    versus frequency $\Omega$ for
    the pure rocking ratchet ---without feedback flashing,
    i.e., $\alpha(t)=1$.--- We have used the ratchet potential of
    Fig.~\ref{fig:pot}. The one-sigma error bars are much smaller than the symbol
    size. Units: $k_BT=1$, $L=1$, $\gamma=1$.
}
\label{fig:output45}
\end{figure}

\begin{figure}
  \begin{center}
    \subfigure[]{\includegraphics[scale=0.6]{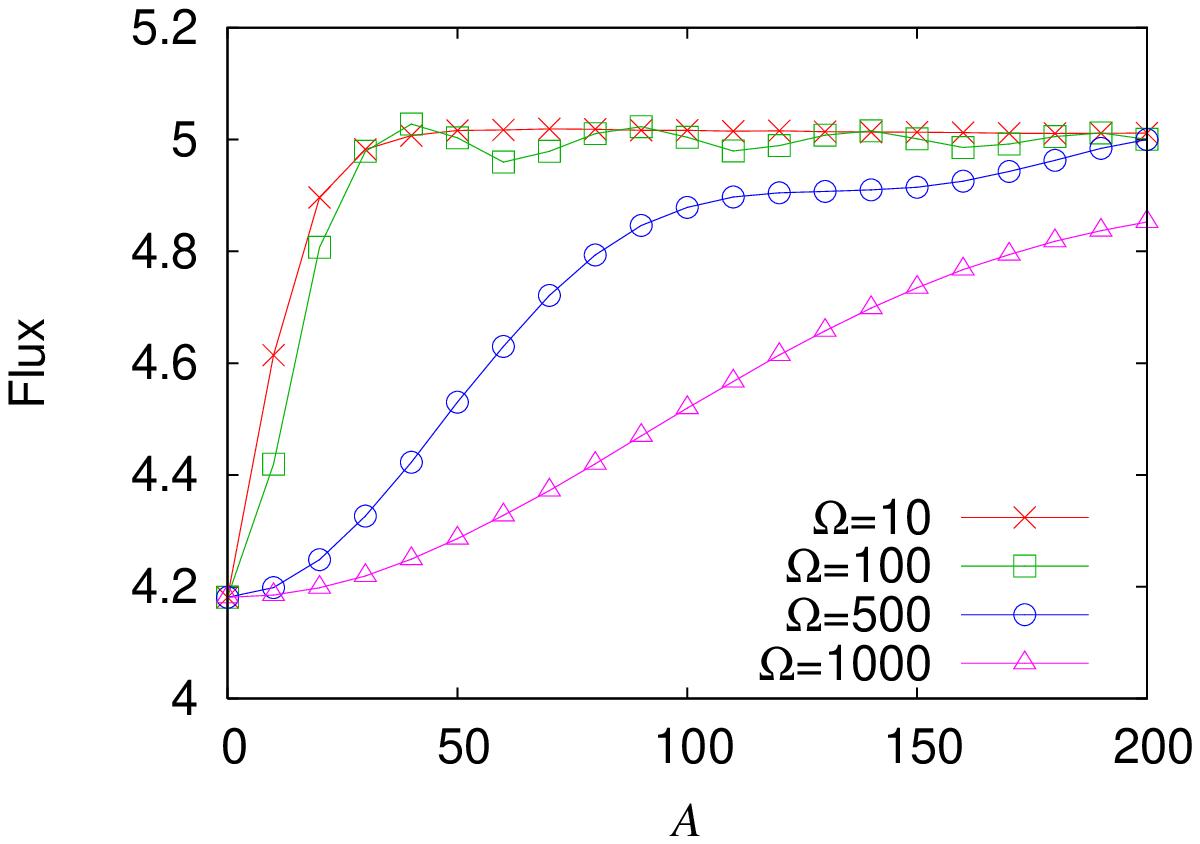}}
    \subfigure[]{\includegraphics[scale=0.6]{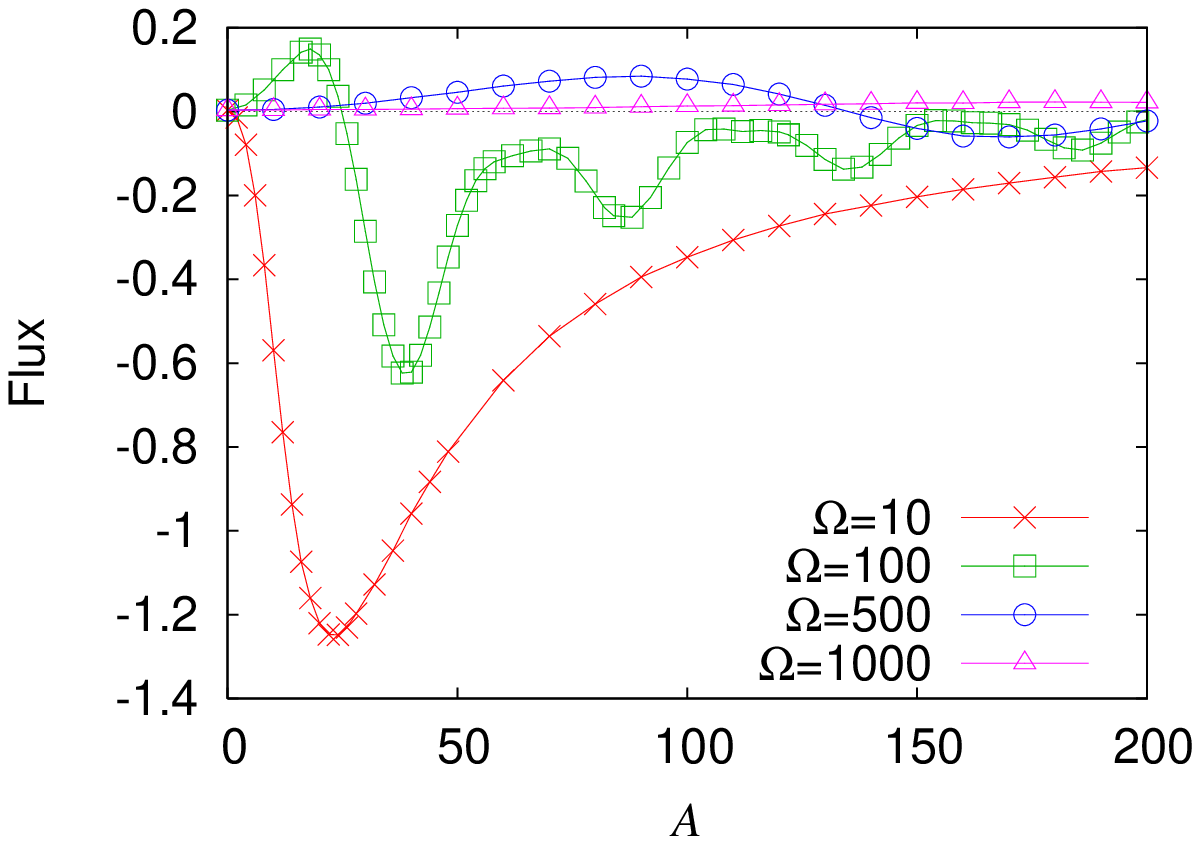}}
\end{center}
\caption{
(Color online) One-particle case.
\emph{Panel (a):} Flux $ \langle \dot x \rangle $ versus amplitude $A$
    for the rocked feedback ratchet. The pure feedback
    ratchet ---without rocking--- corresponds to the point $A=0$.
\emph{Panel (b):} Flux $ \langle \dot x \rangle $ versus amplitude $A$
    for the pure rocking ratchet ---without feedback flashing,
    i.e., $\alpha=1$.--- We have used the ratchet potential of
    Fig.~\ref{fig:pot}. The one-sigma error bars are much smaller than the symbol
    size. Units: $ k_B T = 1 $, $L=1$, $\gamma=1$.
}
\label{fig:flux_vs_A}
\end{figure}

Let us now discuss the case of nonzero temperature. A finite
thermal noise leads to particle diffusion, which provides another
mechanism to overcome the flat regions. This diffusion makes that
the quantized step-like structure for the flux is smeared and
finally wiped out. On the other hand, a surprising effect is found
for this case, namely a flux increase when the feedback policy and
the rocking forcing are both present. Unexpectedly, the resulting
flux is greater than the sum of the flux values due to each
separated effect, as we show in Fig.~\ref{fig:output45}. In fact,
the synchronization of the driving force with the feedback
mechanism gives positive large fluxes even for the case of
negative fluxes for the pure rocking, i.e., with the ratchet
potential always on [compare, for instance, curves for $A=40$ in
panels (a) and (b) of Fig.~\ref{fig:output45}, or curves for
$\Omega=100$ in panels of Fig.~\ref{fig:flux_vs_A}]. Therefore,
adding an external fluctuating force to the maximization of the
instant velocity feedback protocol allows us to improve the
performance of the system in a nontrivial way which to our knowledge has not
been previously reported.
This fact is not only relevant from a theoretical point of view, but also for
experimental ratchet devices designed to maximize the
flux~\cite{cra08,lop08}.
\par

\begin{figure}
    \includegraphics [scale=0.6] {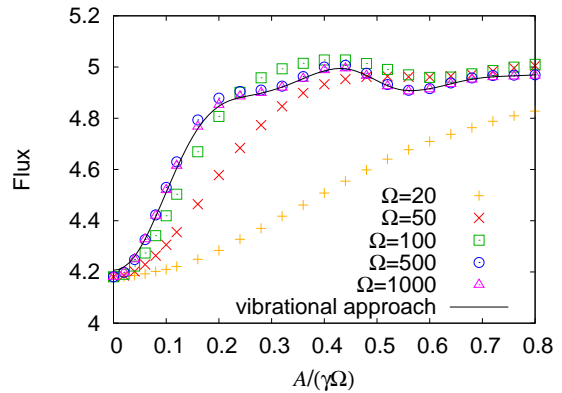}
    \caption{(Color online) Flux $ \langle \dot x \rangle $ versus
      $r=\frac{A}{\gamma\Omega}$ for the one-particle rocked feedback
      ratchet. For increasing frequencies
      the values of the flux tend to the curve of the vibrational
      approximation. We have used ratchet potential~\eqref{smoothpot} with
      $V_0=5$. The one-sigma error bars are much smaller than the symbol
      size. Units: $ k_B T = 1 $, $L=1$, $\gamma=1$.
}\label{fig:flux_VR_article}
\end{figure}

Further insight on the behavior observed in panel (a) of
Figs.~\ref{fig:output45} and~\ref{fig:flux_vs_A} can be obtained studying the
fast driving regime.  In this regime, it is useful to introduce a slow
variable $y(t)$ such that the position $x(t)$ can be written as
$x(t)=y(t)+\psi(t)$, where $\psi(t) = r \sin(\Omega t) $ is the fast
contribution due to the fast driving, and $ r := A/(\gamma \Omega) $. When the
driving is fast enough, a large number of oscillations in $ \psi(t) $ take
place before a significant change in $y(t)$ occurs; thus we can proceed to the
adiabatic elimination of the fast variable $\psi(t)$ by averaging it over
time. This procedure leads to an effective equation for the slow variable
\begin{equation}
\gamma \dot y(t) = \barFeff(y(t)) + \xi(t),
\end{equation}
where $\barFeff(y)=-\barVeff^\prime(y)$, with
\begin{equation} \label{barFeff}
\barVeff(y(t)) := \frac{1}{\cal T} \int_0^{\cal T}
\Veff(y(t)+\psi(s)) \;ds.
\end{equation}
This effective potential allows us to give a closed-form expression for the
flux~\cite{rei02},
\begin{equation}\label{closed}
  \langle\dot{x}\rangle=\frac{L k_BT
    \left[1-e^{(\barVeff(L)-\barVeff(0))/k_BT}\right]}{\gamma\int_0^L dx
    \int_x^{x+L} dy   \;e^{(\barVeff(y)-\barVeff(x))/k_BT}}.
\end{equation}
Note that the potential $\barVeff(y(t))$ only depends on the characteristics
of the driving force through the quotient $r=A/(\gamma \Omega)$, and hence the
same is true for the flux obtained within this fast driving regime. This
approach is known as the {\em vibrational mechanics}
scheme~\cite{Bleckman_book}. It has been successfully applied to the
characterization of the so called vibrational resonance in bistable systems,
both in the absence and presence of noise, as well as to the study of harmful
effects (supression of the firing activity) of strong, high-frequency fields
on the response of excitable systems~\cite{polina00}. In the context of
ratchets, it has been used in the study of the effects of high frequency
modulation on the output of Brownian particles moving in periodic
one-dimensional substrates under the action of low-frequency input
signals~\cite{borro_epl}. The results obtained with this vibrational mechanics
procedure are valid when the rocking force has frequencies much larger than
the rest of characteristic frequencies of the system ~\cite{Bleckman_book}.
The average in Eq.~\eqref{barFeff} makes the original potential barriers
appear effectively lowered and flattened, eventually dissappearing as the
ratio $ r $ increases. In particular, in our system, the periodic part of the
one-particle effective potential $\Veff(x)=\Veff^{\text{per}}(x)-V_0x/L$
becomes smoother and smoother as an effect of the averaging process as $r$
growths, and for large $r$ only the linear term survives in the effective
potential, giving a flux value of $V_0/(L\gamma)$. Panel (a) of
Figs.~\ref{fig:output45} and \ref{fig:flux_vs_A} and
Fig.~\ref{fig:flux_VR_article} show how this value is reached for large
amplitudes.  Indeed, this is the largest value of the flux that has been
obtained in a ratchet device without an a priori bias; see
Fig.~\ref{fig:compara}.

The previous analysis also provides predictions on the dependency
of the flux with the amplitude and frequency of the driving force.
Within the vibrational regime, if the frequency is increased for a
fixed amplitude, i.e., $r$ is decreased, then the flux will
decrease until the value of the pure feedback ratchet [see panel (a) of
Figs.~\ref{fig:output45} and~\ref{fig:flux_VR_article}]. Note that the values
of the flux 
corresponding to low frequencies can be explained with the
adiabatic description in Eq.~\eqref{adiabatic}. On the other hand,
if the amplitude is increased for a fixed frequency, i.e., $r$ is
increased, then the flux will increase from the value of the pure
feedback ratchet up to the maximum value $V_0/(L\gamma)$ [see panel (a) of
Figs.~\ref{fig:flux_vs_A} and~\ref{fig:flux_VR_article}].

The effective potential $\barVeff$ has allowed us to describe the
dynamics in the vibrational regime. We have compared the results
obtained directly from Eqs.~\eqref{barFeff} and~\eqref{closed}
with numerical simulations of the Langevin equation~\eqref{langevin}, with a
good agreement for the fast driving regime
(Fig.~\ref{fig:flux_VR_article}). This stress again the
significance of the vibrational approach that has been revealed as
a useful approach for both qualitative and quantitative
predictions. On the other hand, we have found that fluxes greater
than $V_0/(L\gamma)$ can be attained outside the vibrational
regime in the quasideterministic regime, i.e., for large values of
the potential height and the driving force amplitude. [For example,
ratchet potential~\eqref{smoothpot} with $V_0=40$, and a rocking force
with $A=160$ and $\Omega=290$ gives
$\langle\dot{x}\rangle\simeq 43$ in units $k_BT=1$, $L=1$, and
$\gamma=1$]. This result is in accordance with the results found in
Ref.~\cite{reg02} for a tilted rocking ratchet in the
quasideterministic regime.

\section{Collective ratchet}\label{sec:collective}

The dynamics of the collective ratchet compounded of more than one particle
differs significatively from that of the one-particle ratchet discussed
before. For collective closed-loop ratchets the feedback effectively couples
the particles with each other and no simplifying description in terms of an
effective potential has been found.
\par

The behavior of the deterministic (zero temperature) collective
ratchet is similar to that of the one-particle ratchet, including
the quantization of the flux and the step-like structure commented
in Sec.~\ref{sec:one}.  We shall now focus in the nonzero
temperature case for few- and many-particle collective ratchets
where important differences emerge.
\par

\begin{figure}
    \includegraphics [scale=0.6] {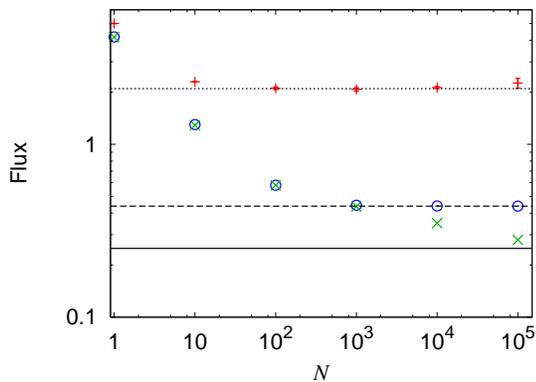}
    \caption{(Color online) Maximum center-of-mass flux versus number of
      particles $N$ for the
optimal rocked feedback ratchet (red $+$), the optimal threshold
protocol (blue $\circ$), the instant maximization protocol ---pure
feedback--- (green $\times$), the optimal rocked flashing ratchet
(dotted line), the optimal periodic protocol (dashed line), and
the optimal rocking ratchet ---pure rocking--- (solid line). We
have used ratchet potential~\eqref{smoothpot} with $V_0=5$.
Where not shown, the error bars are smaller than the
symbol size. Units: $ k_B T = 1 $, $L=1$,
$\gamma=1$.}\label{fig:compara}
\end{figure}

In the few-particle case the maximum averaged center-of-mass flux
is achieved for finite amplitudes and frequencies of the rocking
force. Contrary to the one-particle case, the flux diminishes as
the amplitude increases over its optimal value. On the other hand,
we point out that for collective ratchets the maximum flux
diminishes with the number of particles $N$. For a critical number
of particles the dependence of the flux with $N$ practically
disappears, indicating the transition to the many-particle case;
see Fig.~\ref{fig:compara}. The value for this $N$-independent
maximum flux that is obtained in the many-particle case coincides
with the maximum flux obtained in the corresponding rocked
flashing ratchet (open loop). This coincidence is analogous to the
coincidence between the maximum flux for the threshold protocol in
the many-particle case and the maximum flux obtained from the
corresponding flashing ratchet~\cite{fei06} (see also
Fig.~\ref{fig:compara}). Both of these coincidences can be
interpreted as a consequence of the fact that these feedback
protocols only use one bit of information about the system. This
fact together with the increase of degrees of freedom of the
system as $N$ increases, makes that the relative strength of these
feedback protocols is weakened as the number of particles $N$
increases, and that for systems with a large number of particles
those feedback protocols cannot significantly beat their open-loop
counterparts. In the following we discuss the interesting
cooperative effects appearing in the many-particle case.
\par

In the many-particle case, the force per particle due to the
ratchet potential, $f(t)$ (defined in Sec.~\ref{sec:model}), has a
quasideterministic evolution, as fluctuations in $f(t)$ are
subdominant. The analysis of $f(t)$ has revealed to be very
helpful to understand the dynamics. For the pure feedback ratchet
(without rocking) with many particles the system dynamics gets
trapped with the potential `on' or `off' because the force
fluctuations responsible of the switchings are
negligible~\cite{cao04}. Consequently, the system dynamics is near
equilibrium most of the time and the net force is nearly zero.
This implies an average asymptotic center-of-mass velocity
$\langle\dot{x}_{CM}\rangle$ tending to zero as $N$
increases~\cite{cao04}. However, the introduction of the driving
force allows the system to avoid this trapping and can result in
an increase of the flux.
\par

\begin{figure}
\begin{center}
\subfigure[]{\includegraphics[scale=0.6]{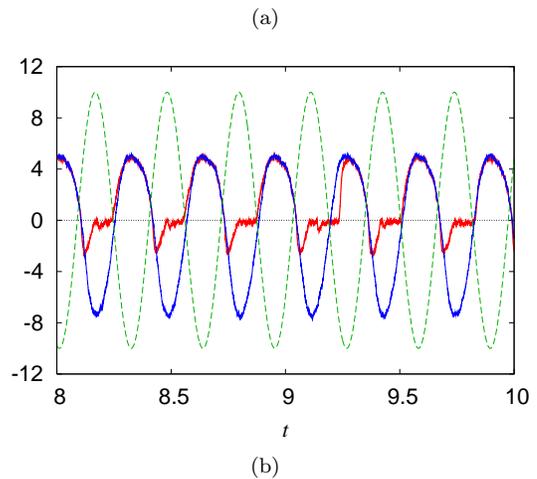}}
\subfigure[]{\includegraphics[scale=0.66]{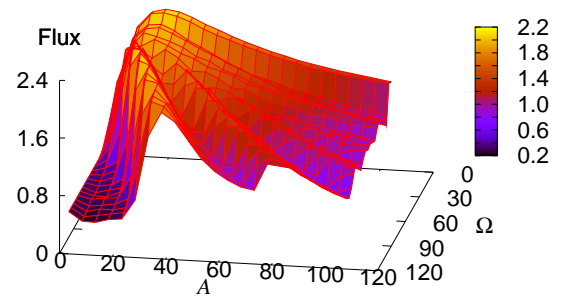}}
\end{center}
\caption{(Color online) Low and medium frequency rocking in the many-particle
  case ($N=10^4$).  \emph{Panel (a):} Time evolution of the force $f$
  [Eq.~\eqref{netforce}] for the rocked feedback ratchet (thick red line) and
  for the pure rocking ratchet (thin blue line), both with a rocking force
  (dashed line) of amplitude $A=10$ and frequency $\Omega=20$.
  \emph{Panel (b):} Flux $ \langle \dot x_{CM} \rangle $ versus amplitude and
  frequency for the rocked feedback ratchet.  We have used ratchet
  potential~\eqref{smoothpot} with $V_0=5$. Units: $k_BT=1$, $L=1$,
  $\gamma=1$.  }
\label{fig:output2}
\end{figure}

Let us first discuss the cases of frequencies $\Omega$ of lower or
similar order to $2\pi/{\cal T}_f$, with ${\cal T}_f$ the
quasiperiod of $f(t)$ for the pure feedback ratchet~\cite{cao04}.
The maximum values of the flux in the many-particle case are
obtained in this low-frequency regime. When the driving force is
added, a complex synchronization appears between the
quasideterministic dynamics of $f(t)$ and the driving force
$A\cos(\Omega t)$. We show in Fig.~\ref{fig:output2} [panel (a)] a
typical time evolution of the forces for this case. The value of
the flux depends on the details of this synchronization and it
shows local maxima and minima when the system's parameters are
tuned. See panel (b) of Fig.~\ref{fig:output2}, where this complex
behavior is shown by computing the flux for a two dimensional grid
of $22\times 20$ points in the $A-\Omega$ plane. For the ratchet
of this Figure the maximum flux $\langle\dot{x}_{CM}\rangle \simeq
2.1$ is achieved for a driving force of amplitude $A\simeq 20$ and
frequency $\Omega\simeq 55$, expressed in units $k_BT=1$, $L=1$,
$\gamma=1$. We want to call the attention to the fact that this
frequency coincides with the characteristic frequency of the
optimal threshold protocol \cite{fei06} and of the optimal
flashing ratchet $\Omega = 2\pi/0.11 \simeq 57$. The maximum flux
for the many-particle rocked feedback ratchet is reached when the
rocking force has this characteristic frequency and pushes forward
during the off period and backward during the on period. This
makes that the ratchet potential only hinders the backward pushes
of the rocking force. Contrary to the one-particle case, the flux
diminishes as the amplitude increases over its optimal value;
compare Fig.~\ref{fig:flux_vs_A}, panel (a), and
Fig.~\ref{fig:output2}, panel (b).
\par

\begin{figure}
    \subfigure[]{\includegraphics [scale=0.6] {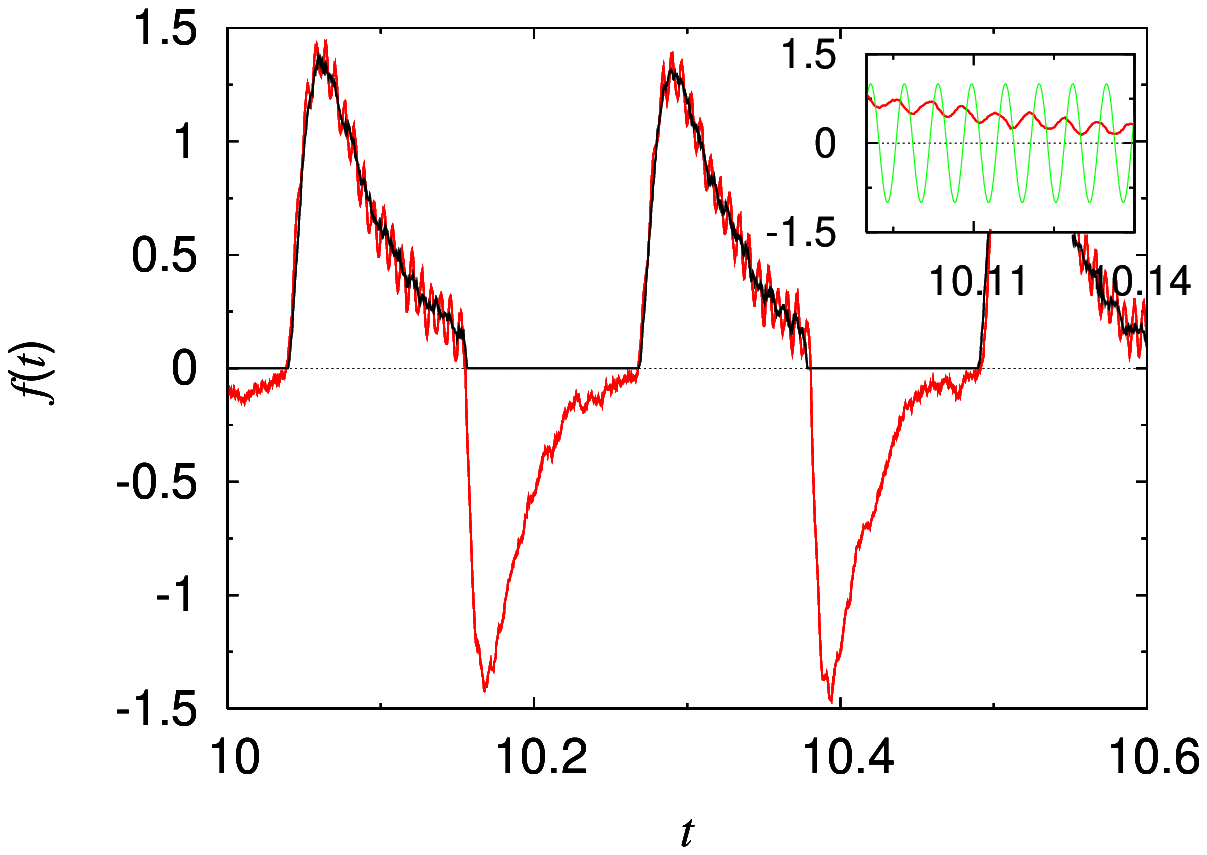}}
    \subfigure[]{\includegraphics[scale=0.6]{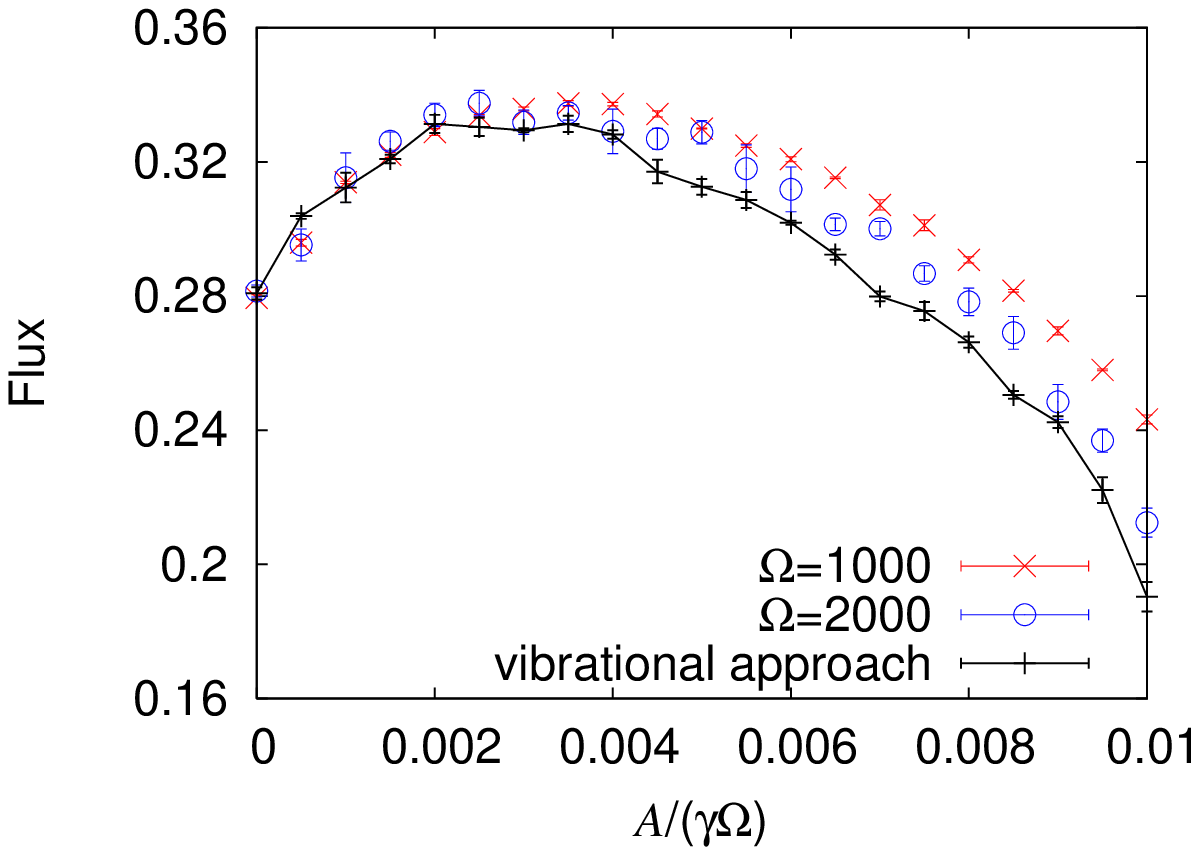}}
    \caption{(Color online) Many-particle case ($N=10^5$) high-frequency
      rocking ($\Omega=1000$) for ratchet potential~\eqref{smoothpot} with
      $V_0=5$.
\emph{Panel (a):} Evolution of the force $f$ [Eq.~\eqref{netforce}]
  for the feedback ratchet rocked with a high-frequency rocking force
  of amplitude $A=1$ compared with the average force
  $ \sum_{i=1}^N \bar F_i/N $ with $ \bar F_i $ given by Eq.~\eqref{barFi}.
  We illustrate in the inset the modulation of $f(t)$
  (thick red line) due to the high-frequency rocking (thin green line).
\emph{Panel (b):} Flux $ \langle \dot x_{CM} \rangle $ versus
  $A/(\gamma \Omega)$ for two high-frequency rockings compared with the
  prediction of the vibrational approximation
  [Eqs.~\eqref{many-vib-1},~\eqref{barFi}]. One-sigma error bars
  are shown. Units: $k_B T=1$, $L=1$, $\gamma=1$.
}
    \label{fig:evol_many_bis}
\end{figure}

On the other hand, in the regime of large frequencies ($\Omega\gg 2\pi/{\cal
  T}_f$) the pattern of $f(t)$ resembles the pattern for the pure feedback
ratchet~\cite{cao04}, but modulated by the high-frequency signal
(Fig.~\ref{fig:evol_many_bis}). For moderate values of the amplitude of the
rocking, the system behaves more or less as if the fluctuations were increased.
Therefore, an enlargement of the flux is possible for appropriate amplitude of
the driving force that succeeds in preventing the trapping similarly to the
so-called threshold protocol~\cite{fei06}. We show in
Fig.~\ref{fig:evol_many_bis} this resonantlike effect for this regime. We note
that for small amplitudes $A$ the system is not able to avoid trapping,
while for too large amplitudes the characteristic quasideterministic $f(t)$
pattern is erased and the flux goes to zero.

As in the one-particle ratchet, a vibrational regime appears when the
displacements induced by the driving force are faster than the effects of the
other terms. This is shown in panel (b) of
Fig.~\ref{fig:evol_many_bis}. In this panel, the dependence of the flux on the
ratio $r=A/(\gamma \Omega)$ for two different high-frequency rocking forces is
compared with the flux obtained by assuming an effective dynamics defined as
follows. As for the case $N=1$, we introduce the slow variables $y_i(t) =
x_i(t) - \psi(t)$, with $\psi(t) = r \sin(\Omega t)$ the displacements induced
by the fast driving. Numerical simulations confirm that the dynamics in this
regime is governed by the slow variables verifying the following averaged
evolution equations:
\begin{equation}\label{many-vib-1}
\gamma \dot y_i(t) = \bar F_i(y_1(t),\ldots,y_N(t))+\xi_i(t),
\end{equation}
where
\begin{equation}\label{barFi}
\begin{split}
\bar F_i(y_1, \ldots,y_N) = \frac{1}{\cal T} \int_0^{\cal T}
\! ds \; \alpha(y_1+\psi(s), \ldots,y_N+\psi(s)) \\
\times F_i(y_1+\psi(s), \ldots,y_N+\psi(s)),
\end{split}
\end{equation}
with $\alpha$ given by Eq.~\eqref{alfa}. This implies as before
that within this regime the flux only depends on the
characteristics of the driving force through the quotient
$r=A/(\gamma\Omega)$. We have numerically checked this for the few
and the many-particle cases with high-frequency driving forces,
finding a better agreement in the few-particle case. However, we
have also found a good agreement in the many-particle case for
small values of the rocking amplitude [see
Fig.~\ref{fig:evol_many_bis} (b)] when $\sum_{i=1}^N \bar F_i/N$ is a good
average description of the force $f(t)$ [see
Fig.~\ref{fig:evol_many_bis} (a)]. In addition to computations
with ratchet potential \eqref{smoothpot}, we have also
performed computations with other potentials and found analogous
results.
\par

\section{Concluding remarks}\label{sec:conclusions}
In this paper we have studied the effects of rocking a feedback ratchet.
The interplay between the rocking and the feedback policy gives an
intriguing rich dynamics that we have analyzed and discussed.
\par

For the one-particle rocked feedback ratchet we have found an effective
description in terms of a tilted rocking ratchet. Our simulations for
rocked feedback ratchets show a relevant effect, namely, the magnification
of the flux with respect to both the pure rocking and the pure feedback.
That is, the rocked feedback ratchet is able to give fluxes even larger
than the sum of the two fluxes separately.
At this point, we remark that one of the main advantages of feedback ratchets
over their open-loop counterparts is their ability to enlarge the particle
flux, as it has been proved theoretically~\cite{cao04} and
experimentally~\cite{lop08}. In that sense, the introduction of the
fluctuating force in feedback ratchets provides a way to further enhance the
flux performance. In fact, the one-particle rocked feedback ratchet studied
here gives the maximum flux that has been achieved in a ratchet device without
an a priori bias (see Fig.~\ref{fig:compara}). This improvement in the flux
performance is relevant for nanotechnological applications of the ratchets. In
addition, the observed dependence of the flux on the frequency and amplitude
of the rocking signal has been explained for the whole range of
parameters of interest.
\par

The rocking term also helps to enlarge the flux in the few- and
many-particle case, as we have shown in Fig.~\ref{fig:compara}. In
this respect we highlight that the increase of the number of
particles effectively decreases the strength of the feedback in
the control, and in the limit of infinite number of particles this
closed-loop protocol cannot give fluxes greater than its
corresponding open-loop protocols. We have numerically shown the
dependence of the flux with the amplitude and frequency of the
driving force for these rocked feedback collective ratchets. The
details of this dependence follow from the synchronization between
the driving force and the feedback. In addition, we have found a
new resonantlike effect when the amplitude of the rocking is tuned
in the regime of  high-frequency signals. This later effect can be
viewed as similar to an effective enlargement of the fluctuations
in the net force, which prevents the trapping of the dynamics near
equilibrium and results in an increase in the flux.

To sum up, we have proposed and discussed a new closed-loop
ratchet that is able to perform better than other known ratchets
as a consequence of the nontrivial interplay of the feedback
scheme and the rocking force. We have found an effective potential for the
one-particle ratchet that explains the effective bias of the system and the
a priori unexpected high values of the flux; it also has allow us to provide a
closed expression for the flux in the vibrational regime. We have
also analyzed the rich dynamics for the collective ratchet describing the
vibrational regime and new resonantlike effects.

\begin{acknowledgments}
MF and FJC acknowledge financial support from MCYT (Spain) through the Research
Project FIS2006-05895, from the ESF Programme STOCHDYN, and from
UCM and CM (Spain) through CCG07-UCM/ESP-2925 and CCG08-UCM/ESP-4062. JPB
acknowledges support from
the Universidad de Huelva (project FQM-276) and from the Ministerio de
Ciencia e Innovaci\'on (Spain) under the research project FIS2008-02873.
\end{acknowledgments}


\end{document}